\documentclass[a4paper]{article}

\usepackage{amsmath,graphicx}
\usepackage{amsfonts}
\usepackage{url}
\usepackage{tabularx}
\usepackage{siunitx}
\usepackage{multirow}
\usepackage{subfigmat}
\usepackage{mathtools}
\usepackage{color}

\usepackage[linesnumbered,ruled,vlined,commentsnumbered]{algorithm2e}

\renewcommand{\v}[1]{\ensuremath{\mathbf{#1}}}

\usepackage{INTERSPEECH2021}

\title{Semi-Supervised Training with Pseudo-Labeling for\\ End-to-End Neural Diarization}
\name{Yuki Takashima$^1$, Yusuke Fujita$^1$, Shota Horiguchi$^1$, Shinji Watanabe$^{2,3}$, Paola Garc\' ia$^3$, Kenji~Nagamatsu$^1$}
\address{
$^1$ Hitachi, Ltd. Research \& Development Group, Japan\\
$^2$ Language Technologies Institute, Carnegie Mellon University, USA\\
$^3$ Center for Language and Speech Processing, Johns Hopkins University, USA}
\email{yuki.takashima.ot@hitachi.com}

\begin{document}

\maketitle
\begin{abstract}
In this paper, we present a semi-supervised training technique using pseudo-labeling for end-to-end neural diarization (EEND).
The EEND system has shown promising performance compared with traditional clustering-based methods, especially in the case of overlapping speech.
However, to get a well-tuned model, EEND requires labeled data for all the joint speech activities of every speaker at each time frame in a recording.
In this paper, we explore a pseudo-labeling approach that employs unlabeled data.
First, we propose an iterative pseudo-label method for EEND, which trains the model using unlabeled data of a target condition.
Then, we also propose a committee-based training method to improve the performance of EEND.
To evaluate our proposed method, we conduct the experiments of model adaptation using labeled and unlabeled data.
Experimental results on the CALLHOME dataset show that our proposed pseudo-label achieved a 37.4\% relative diarization error rate reduction compared to a seed model.
Moreover, we analyzed the results of semi-supervised adaptation with pseudo-labeling.
We also show the effectiveness of our approach on the third DIHARD dataset.
\end{abstract}

\section{Introduction}
\label{sec:intro}

Speaker diarization is the process of partitioning a speech recording into homogeneous segments associated with each speaker.
This process is an essential part of multi-speaker audio applications such as generating transcriptions from meetings~\cite{Tranter2006,Anguera2012}.
Recent studies~\cite{conf/asru/ZorilaBDH19,conf/asru/KandaHFXNW19} have shown that accurate diarization improves the performance of automatic speech recognition~(ASR).

The traditional speaker diarization approach uses a clustering-based method that relies on multiple steps: voice activity detection (VAD), speech segmentation, feature extraction, and clustering.
VAD is the process of filtering out the non-speech parts from an input speech.
Speech regions are then split into multiple speaker-homogeneous segments, and frame-level speaker embeddings are extracted.
The recent progress on deep learning has made it possible to compute better speaker representation such as x-vectors~\cite{conf/icassp/SnyderGSPK18} and d-vectors~\cite{conf/icassp/WanWPL18}.
Once the embeddings are obtained, a clustering method is needed.
The commonly used methods are agglomerative hierarchical clustering~\cite{conf/slt/SellG14}, k-means clustering~\cite{conf/interspeech/DimitriadisF17}, spectral clustering~\cite{conf/icassp/WangDWMM18,conf/interspeech/LinYLBB19}, and affinity propagation~\cite{Frey07AffinityPropagation}.
Recently, neural network-based clustering has been explored~\cite{li2019discriminative}.
Although clustering-based methods performed well, they are not optimized to directly minimize diarization errors because clustering is an unsupervised method.
To directly minimize diarization errors in a supervised manner, clustering-free methods have been studied~\cite{conf/icassp/ZhangWZP019,conf/interspeech/FujitaKHNW19,conf/asru/FujitaKHXNW19}.

End-to-end neural diarization~(EEND)~\cite{conf/interspeech/FujitaKHNW19,conf/asru/FujitaKHXNW19} is a promising direction for speaker diarization.
EEND uses a single neural network that maps a multi-speaker audio to joint speech activities of multiple speakers.
In contrast to most of the clustering-based methods, EEND handles overlapping speech without using any external module.
Inspired by its successful results, several variants of EEND have been proposed~\cite{journals/corr/abs-2006-01796,horiguchi2020end,xue2020online,journals/corr/abs-2011-02678,9383555}.
However, to get a well-tuned model, these methods require labeled data for all the joint speech activities of every speaker at each time frame in a recording.
It is difficult to collect such data in real environments, especially in case there are many speaker overlaps.

Therefore, in this paper, we investigate a training strategy using unlabeled data for EEND.
To achieve this objective, we employ
pseudo-labeling~\cite{pseudo2013label,conf/icassp/Kahn0H20} that was originally introduced as a semi-supervised learning algorithm to train a model with a limited amount of labeled data and a large amount of unlabeled data.
In this work, we assume an adaptation scenario of EEND where we have a seed model trained on labeled data of a source domain.
First, we explore an iterative pseudo-label method for fine-tuning an EEND model using only unlabeled data of a target condition.
To improve the quality of the pseudo-label, we alternately execute pseudo-labeling and fine-tuning the model with the generated pseudo-label, which gradually refines the pseudo-label.
Moreover, we propose a committee-based training method.
In the committee-based learning~\cite{conf/interspeech/KandaHLK16}, multiple recognizers are trained, and then the obtained results are used for data selection.
Inspired by this approach, we train multiple diarization models, then the pseudo-labels generated from them are combined.
This composite pseudo-label is used for adaptation.
Through the experiments on CALLHOME~\cite{callhome}, we demonstrate that our method achieves a 37.4\% relative diarization error rate (DER) reduction compared to the seed model.
Our proposed method with unlabeled data also outperformed the well-adapted model with only labeled data.

Lastly, we conduct the experiments on the third DIHARD dataset~\cite{ryant2020third} to investigate the potential of semi-supervised training for challenging speaker diarization scenarios.
To get a more accurate pseudo-label, we combine the results of clustering-based methods and EEND-based methods.
We perform the system combination by using DOVER-Lap~\cite{journals/corr/abs-2011-01997} that is a voting algorithm for speaker diarization.
By analyzing the results~\cite{horiguchi2021hitachi}, we show that this approach improves the performance of not only EEND itself but also the final fusion.
Thanks to this technique, we also achieved the second-best in the DIHARD III challenge.

\section{Related work}
\label{sec:related}
Neural network-based speaker diarization~\cite{conf/interspeech/FujitaKHNW19,conf/asru/FujitaKHXNW19,conf/icassp/KinoshitaDAN20,medennikov2020target} has gained a lot of attention due to its simplicity in training and inference.
In this work, we focus on EEND that shows substantial progress with various extensions.
In~\cite{journals/corr/abs-2006-01796,horiguchi2020end}, the authors show experimental analyses and proposals for increasing the number of speakers.
The EEND models are also evaluated in online scenarios~\cite{xue2020online,journals/corr/abs-2011-02678}.
However, these approaches require labeled data for training the model.
In this paper, we investigate a semi-supervised training strategy to effectively utilize unlabeled data for EEND.

The pseudo-label~\cite{pseudo2013label} is a label generated from a pre-trained model with labeled data, that can be used for semi-supervised and unsupervised training with unlabeled data.
In ASR, it is well-known that this technique improves recognition performance when a large amount of untranscribed data is available~\cite{LAMEL2002115,conf/interspeech/XuSSL14}.
Because the performance improvement depends on the seed model, there are several investigations on getting a high reliable label~\cite{conf/icassp/Charlet01,conf/asru/VeselyHB13}.
Most recently, Kahn~{\it et al.}~\cite{conf/icassp/Kahn0H20} investigated end-to-end ASR with pseudo-labeling, and Xu~{\it et al.}~\cite{conf/interspeech/XuLKHSC20} proposed iterative pseudo-labeling as an extension of it.
For speaker recognition, Cai~{\it et al.}~\cite{cai2020iterative} proposed an iterative framework with pseudo-labeling to train a speaker embedding network.
While these studies focus on utilizing a large amount of unlabeled data, we aim to adapt the model to a target condition using unlabeled data.

The committee-based learning trains multiple recognizers, and uses the outputs obtained from them for selective-sampling problem from unlabeled data.
Hamanaka~{\it et al.}~\cite{conf/icassp/HamanakaSFEK10} applied this framework to speech recognition.
Kanda~{\it et al.}~\cite{conf/interspeech/KandaHLK16} investigated the use of heterogeneous neural networks as committee members for semi-supervised acoustic model training.
Huang~{\it et al.}~\cite{conf/interspeech/HuangYGL13} proposed a multi-system combination and confidence re-calibration approach to improve the transcription inference and data selection.
While these methods focused on data selection for unlabeled data, we apply this idea to obtain the composite output via the committee.

\section{End-to-end neural diarization: Review}
Given a $T$-length time sequence of acoustic features as an input, EEND processes it using bi-directional long short-term memory (BLSTM) or Transformer encoders to obtain an embedding.
Then, a decoder network (e.g. a linear transformation or LSTM) with a sigmoid activation is applied to calculate an output probability $\hat{\v{y}}_t\in (0,1)^S$ for the number of speakers $S$ at time $t$.
In the training phase, the EEND is optimized using the permutation-free training scheme~\cite{conf/icassp/YuKT017,conf/icassp/HersheyCRW16}, {\it i.e.}, the loss is calculated between the neural network output $\hat{\v{y}}_{t}$ and reference label $\v{y}_{t}\in \{0,1\}^S$ as follows:
\begin{align}
\label{eq.pit}
L_{\mathrm{PIT}} = \frac{1}{ST}\min_{\phi \in \mathrm{perm}(S)}\sum_{t=1}^T \mathrm{BCE}(\hat{\v{y}}_{t}, \v{y}^\phi_{t}),
\end{align}
where $\mathrm{perm}(S)$ is a set of all possible permutations of a sequence $(1, \dots, S)$ and $\mathrm{BCE}(\cdot,\cdot)$ is an element-wise binary cross-entropy function followed by the summation of all elements.
$\v{y}^\phi_{t}$ indicates the reference label at time $t$ after the permutation $\phi$.

The EEND-based system is usually trained using a large amount of simulated data~\cite{conf/interspeech/FujitaKHNW19}, then adapted using real audio mixtures of the target environment.
To do that, we need to prepare a pair of the audio mixture and the corresponding annotation.

\begin{figure}[tb]
    \centering
        \includegraphics[clip,keepaspectratio, scale=0.4]{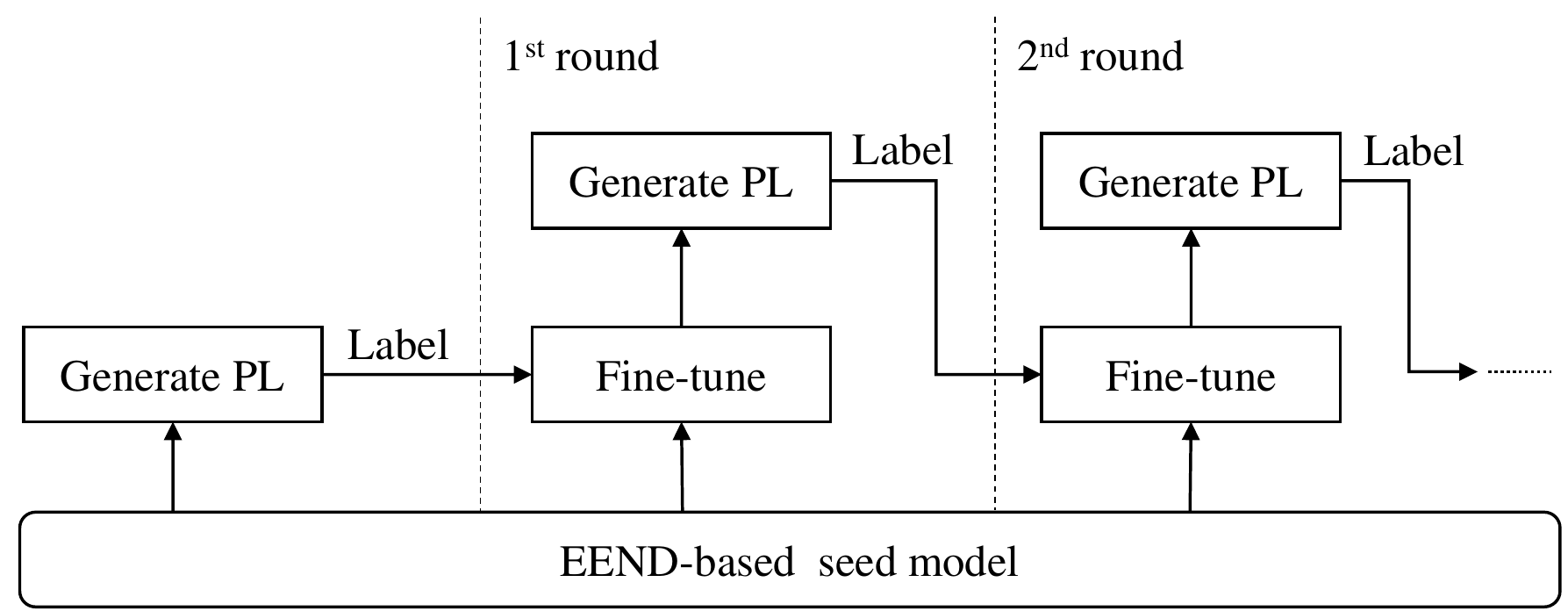}
        \caption{Overview of the iterative pseudo-label method.
        PL indicates the pseudo label.}
        \label{fig:IPL}
\end{figure}

\section{Proposed method}
\label{sec:prop}
In this section, we first introduce the iterative pseudo-label method for EEND using only unlabeled data.
Then, we propose a committee-based training method for speaker diarization.

\subsection{Iterative pseudo-label method for EEND}
\label{sec:IPL}
EEND is usually trained on simulated data, and then adapted to the target condition with labeled data.
When we perform unsupervised adaptation in this scenario, we should consider domain shift that is the difference between the distributions of the source and target domains.
The unexpected performance degradation can be caused by some errors in the pseudo-label.
To alleviate this problem, we propose the iterative pseudo-label method that continuously refines the pseudo-label.

Figure~\ref{fig:IPL} shows a training procedure of our proposed method.
Given an EEND-based seed model trained on labeled data, we calculate the neural network output for unlabeled data to generate the pseudo-label.
Then, we fine-tune the EEND-based seed model using this pseudo-label with a small number of epochs.
The fine-tuned model is used to update the pseudo-label.
We perform this process for fine-tuning the EEND-based seed model and generating the pseudo-labels in each round.
By updating the pseudo-label iteratively, we prevent overfitting caused by undesirable errors.
We note that our proposed method can be applied to any variants of EEND.

\subsection{Committee-based training for speaker diarization}
\label{sec:committee}

\begin{figure}[tb]
    \centering
        \includegraphics[clip,keepaspectratio, width=\linewidth]{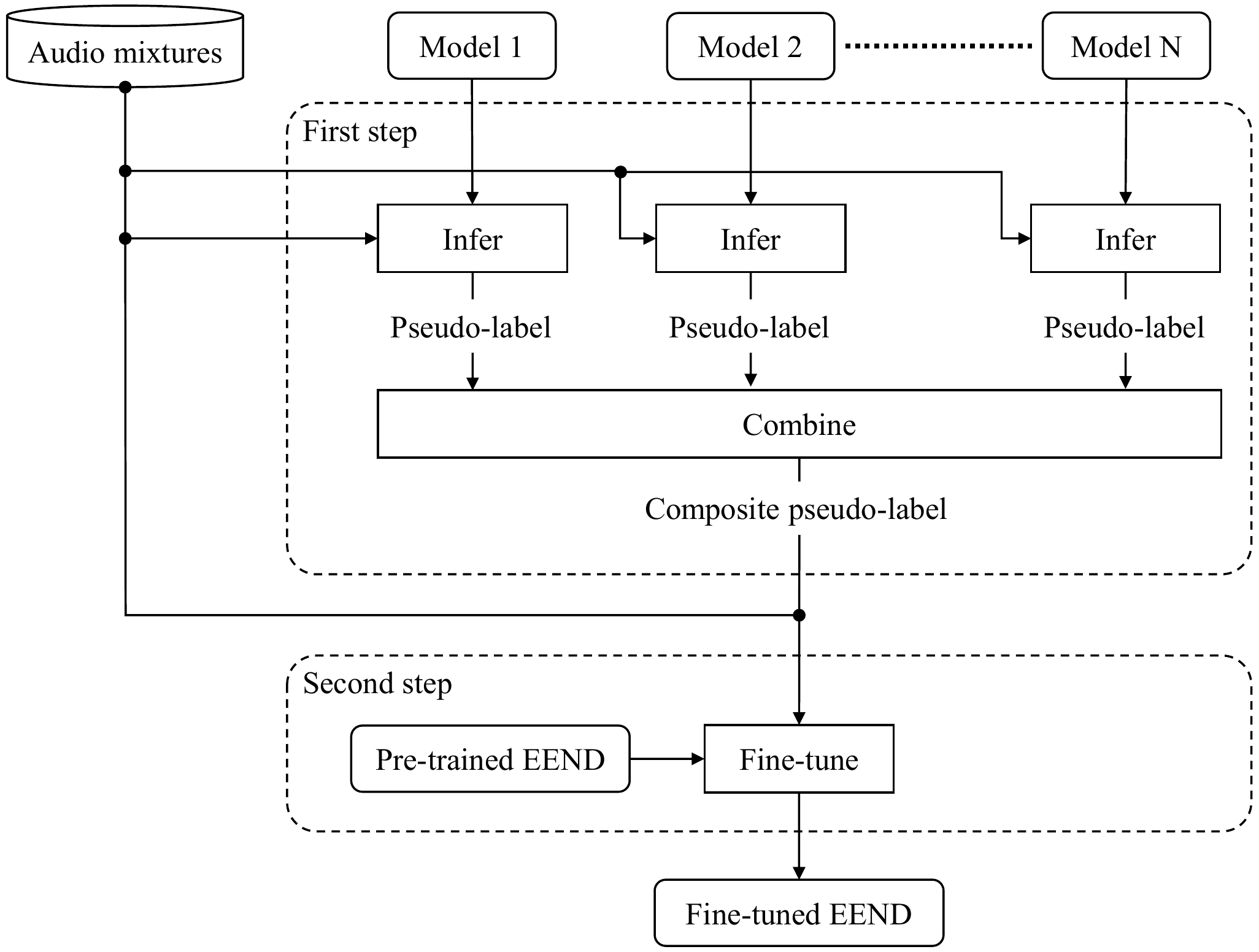}
        \caption{Flow of our proposed committee-based training method.}
        \label{fig:committee}
\end{figure}

Our proposed committee-based training consists of two steps:
pseudo-label combination and fine-tuning using the composite pseudo-label, as depicted in Fig.~\ref{fig:committee}.
In the pseudo-label combination step, given $N$ diarization models as committee members, we generate the pseudo-labels for audio mixtures that are unlabeled data of a target condition.
Note that we can use any type of diarization system as a model.
Then, the generated pseudo-labels are combined to enhance the quality of the pseudo-label.
In the second step, we fine-tune the EEND-based model using audio mixtures with its pseudo-label.

\section{Experimental results}
\label{sec:exp}

\subsection{Conditions}
\label{sec:cond}

\begin{table}[tb]
\vspace{-8pt}
\caption{Statistics of training/adaptation/test sets.
The training set is simulated mixtures.
The adaptation and test sets are the subsets of the CALLHOME dataset.}
\vspace{-3pt}
\label{tbl:set}
\centering
\scalebox{0.9}{
\begin{tabular}{@{}lrrrr@{}} \hline\hline
& Num. & Num. of & Avg. dur. & \multicolumn{1}{c}{Overlap} \\
& spk & mixtures & (sec) & ratio (\%) \\\hline
\:Traning set & 1-4 & 100,000 & 130.0 & 29.7 \\
\:Adaaptation set & 2-7 & 249 & 125.8 & 17.0 \\
\:Test set & 2-6 & 250 & 123.2 & 16.7 \\ \hline\hline
\end{tabular}
}
\vspace{-4pt}
\end{table}

The proposed method was evaluated for variable-speaker audio mixtures.
We prepared a simulated training set based on~\cite{conf/asru/FujitaKHXNW19} to train the seed model.
We also prepared real adaptation/test sets from CALLHOME~\cite{callhome}.
The statistics of the datasets are listed in Table~\ref{tbl:set}.
For the CALLHOME set, we employed identical sets as the ones given in Kaldi CALLHOME\_diarization v2 recipe\footnote{https://github.com/kaldi-asr/kaldi/tree/master/egs/callhome\_diarization}.
During adaptation, we can access reference labels of the adaptation set as well as audio mixtures.
On the other hand, for the test set, we can only access audio mixtures, not reference labels.

In these experiments, we employed SC-EEND~\cite{journals/corr/abs-2006-01796}.
The input features were 23-dimensional log-Mel-filterbanks with a \SI{25}{\ms} frame length and \SI{10}{\ms} frame shift.
For the experiments, each feature was concatenated with those from the previous 14 frames and subsequent 14 frames.
After subsampling the concatenated features by a factor of 20, we used four encoder blocks and with 384 attention units containing six heads.

We used diarization error rate~(DER) as an evaluation metric.
A \SI{250}{\ms} collar was employed at the start and end of each segment. Note that we included errors in overlapped segments and SAD-related errors for the DER calculation, whereas most works~\cite{conf/icassp/ZhangWZP019,conf/interspeech/McCreeSG19} in literature did not evaluate such errors.

\subsection{Evaluation with the iterative pseudo-label method}
\label{sec:eval_IPL}

\begin{figure}[tb]
    \centering
        \includegraphics[clip,keepaspectratio, width=\linewidth]{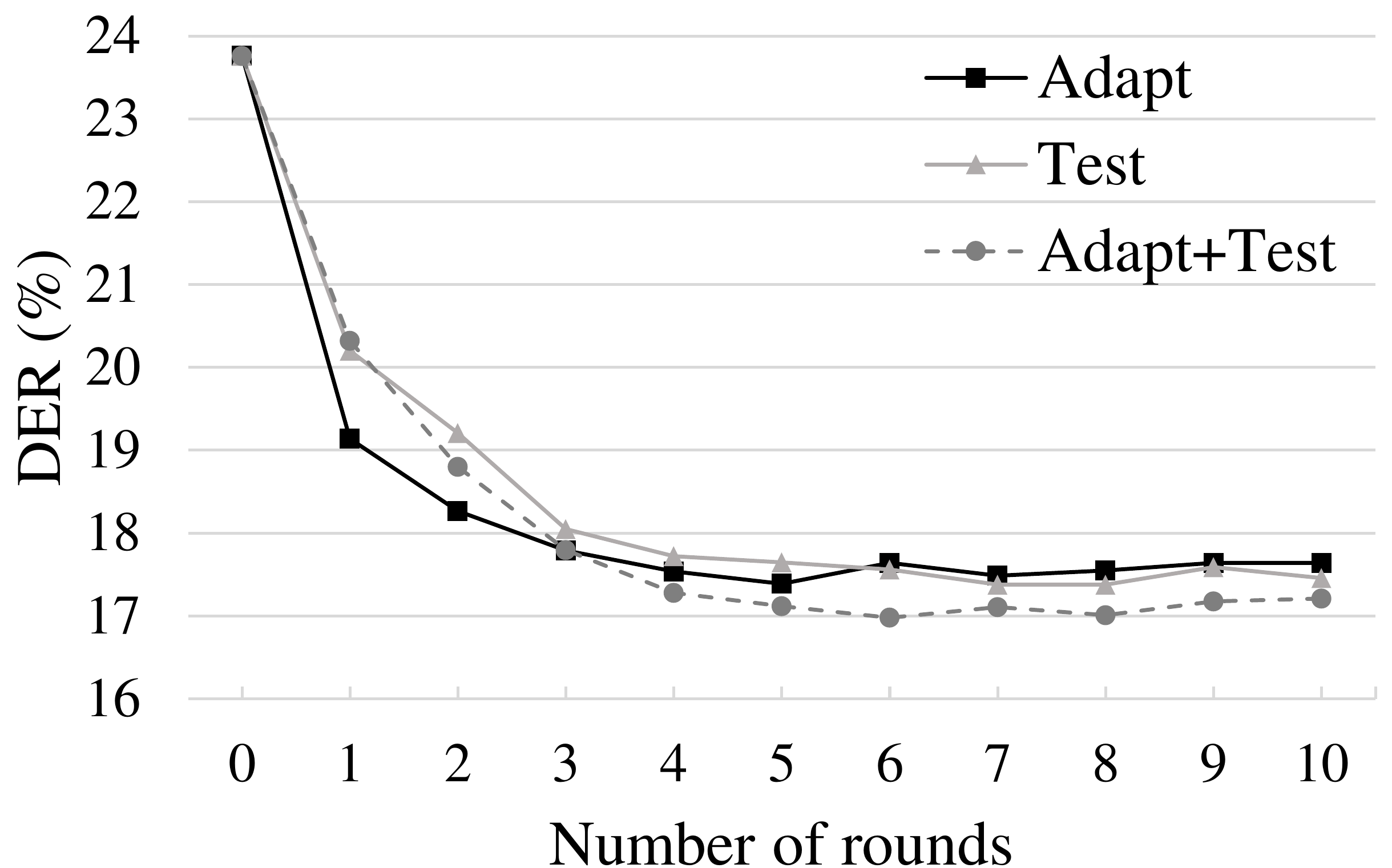}
        \caption{DERs (\%) for unsupervised adaptation with the pseudo-label using different adaptation settings.}
        \label{fig:ua}
\end{figure}

\begin{table}[tb]
    \caption{Detailed DERs (\%) associated with each number of speaker when using both adaptation and test sets for adaptation.}
    \vspace{-3pt}
    \centering
    \setlength{\tabcolsep}{3.0pt}
    \begin{tabular}{c|ccccc} \hline\hline
    & \multicolumn{5}{c}{Number of speakers} \\
    Number of rounds & 2 & 3 & 4 & 5 & 6 \\ \hline
    1 & 14.48 & 18.61 & 25.70 & 36.54 & 37.74 \\
    2 & 12.53 & 17.05 & 25.20 & 34.60 & 35.61 \\
    3 & 11.43 & 16.09 & 24.26 & {\bf 33.17} & {\bf 35.37} \\
    4 & 10.67 & 15.95 & 23.19 & 33.24 & 35.54 \\
    5 & {\bf 10.34} & {\bf 15.75} & {\bf 23.10} & 33.56 & 36.10 \\
    \hline\hline
    \end{tabular}
    \label{tab:detailspk}
\end{table}

In this section, we report the results for adaptation with the iterative pseudo-label method.
We conducted three adaptation settings of unlabeled data: adaptation, test, and both adaptation and test sets.
We note that any label of the CALLHOME dataset was not used even for the adaptation set in this scenario.
Before adaptation, we trained a seed model using the training set listed in Table~\ref{tbl:set}.
When we used either the adaptation set or the test set for adaptation, the number of epochs for each round was 7.
When we used both sets, the number of epochs for each round was 15.

Figure~\ref{fig:ua} shows the results of DERs where Round 0 indicates the result of the seed model.
We can see that iterative processing improves performance consistently.
The adaption using the adaptation set achieved a 26.5\% relative DER reduction compared to the seed model.
In the case of adaptation using the test set, we observed a similar trend; however, the score was slightly worse than that using the adaptation set.
We hypothesized that the reason is the overfitting due to the use of the same data for both adaptation and evaluation.
Because the pseudo-label contains some missing speakers and wrong labels, a careful learning strategy is required (e.g. early stopping).
Finally, by using both the adaptation and test sets for adaptation, we achieved 16.98\% of DER in the sixth round that is the best performance in this figure.
These results show that our proposed method is an effective strategy to adapt the model without any label of the target condition.
Because the DERs became saturated around the fifth round for all adaptation settings, we use the results up to the fifth round in the following experiments.

Table~\ref{tab:detailspk} shows the detailed DER results for each number of speakers.
Here, we analyzed the result (DER of 17.12\%
in Fig.~\ref{fig:ua}) for the fifth round using both adaptation and test sets.
For the cases of less than five speakers, we can see significant improvements as the number of rounds increased.
On the other hand, for five- and six-speaker cases, the degradation of the DER was started from the fourth round.
This result suggests that the overall performance can be inferior by more rounds.
However, it is not easy to argue the case of a higher number of speakers because there are a small number of recordings for more than five speakers in the CALLHOME dataset.
The numbers of recordings for five- and six-speaker mixtures are only five and three, respectively.
We will further investigate the evaluation on other datasets in the future.

\subsection{Committee-based training for speaker diarization}

\begin{table}[tb]
    \caption{DERs (\%) on the CALLHOME dataset. PL indicates the pseudo-label. The first part of the results (from (1) to (5)) is obtained with \textbf{unsupervised} adaptation. The rest of the results (from (6) to (8)) is obtained with \textbf{supervised} adaptation.}
    \vspace{-3pt}
    \centering
    \setlength{\tabcolsep}{3.0pt}
    \begin{tabular}{l|cccccc} \toprule
     \multicolumn{1}{c}{} & & Labeled & \multicolumn{3}{c}{Unlabeled} & \\ \cmidrule(r){4-6}
    & Seed Model & Adapt & Adapt & Test & PL from & DER\\ \hline
    (1)  & \multicolumn{5}{c}{--- Initial model ---} & 23.76 \\\hline
    (2) & (1) & & \checkmark & & (1) & 17.39 \\
    (3) & (1) & & & \checkmark & (1) &  17.65 \\
    (4) & (1) & & \checkmark & \checkmark & (1) & 17.12 \\
    (5) & \multicolumn{5}{c}{--- Committee of (2)(3)(4) ---} & \textbf{16.72}\\\hline\hline
    (6) & (1) & \checkmark & & & & 15.23 \\
    (7) & (6) & & & \checkmark & (6) & 15.47 \\
    (8) & (6) & & & \checkmark & (5) & \textbf{14.88} \\
    \bottomrule
    \end{tabular}
    \label{tbl:committee}
\end{table}

In this section, we investigate the impact of pseudo-labels on committee-based training.
The results are shown in Table~\ref{tbl:committee}.
In the first step shown in Fig.~\ref{fig:committee}, we used the three models ($N=3$) described in Section~\ref{sec:IPL} as committee members.
The system combination was performed using DOVER-Lap~\cite{journals/corr/abs-2011-01997} that is a voting algorithm for speaker diarization.
As a result, the composite pseudo-label was generated in fully-unsupervised manner for the target condition.
In the second step, we prepared a pre-trained SC-EEND using the adaptation set with supervision, then fine-tuned it using the test set with the composite pseudo-label.
The number of epochs for supervised adaptation and committee-based unsupervised adaptation were 100 and 10, respectively.
Therefore, overall, this experimental setting is a semi-supervised adaptation.

As shown in Table~\ref{tbl:committee}, the initial model trained on simulated data achieved 23.76\% of DER.
Systems (2) to (4) correspond to the results of the iterative pseudo-label method in the fifth round of Fig.~\ref{fig:ua}.
System (5) shows the DER of the composite pseudo-label obtained by combining the results of above three systems.
We achieved 29.6\% relative DER reduction compared to the initial model without using any label information of the target condition.
From systems (6) to (8), we demonstrated the evaluations with labeled adaptation data.
The fully supervised adaptation (6) achieved 15.23\% of DER.
Comparing systems (5) and (6), we achieved the competitive result although even if system (5) does not use any labeled data.
For systems (7) and (8), we fine-tune system (6) as a seed model.
Unfortunately, the DER of system (7) was higher than that before adaptation.
This result suggests that undesirable errors of the pseudo-label were dominant compared to valuable information for adaptation.
On the other hand, our proposed committee-based adaptation outperformed the fully-supervised adaptation (6).
These results show that the pseudo-label generated in the fully unsupervised manner has complementary information to the manual annotation.

\subsection{Evaluation with semi-supervised adaptation on DIHARD III}
\label{sec:dihard3}

\begin{table}[tb]
    \caption{Detailed DERs (\%) evaluated on the DIHARD III challenge.
    DER is composed of Misses (MI), False alarms (FA), and Confusion errors (CF).}
    \vspace{-3pt}
    \centering
    \setlength{\tabcolsep}{3.0pt}
    \begin{tabular}{l|c|ccc} \hline\hline
    & & \multicolumn{3}{c}{DER breakdown} \\
    Model & DER & MI & FA & CF \\ \hline
    Supervised EEND-EDA & 19.04 & 8.27 & 4.84 & 5.93 \\
    First fusion & 17.21 & 8.21 & 4.22 & 4.77 \\
    Semi-supervised EEND-EDA & 17.81 & 8.6 & 4.22 & 5.00\\
    Second fusion & 16.94 & 8.32 & 4.20 & 4.42\\
    \hline\hline
    \end{tabular}
    \label{tab:result_DH}
\end{table}

Finally, we introduce semi-supervised adaptation with pseudo-labeling for EEND as an application of our proposed method.
As mentioned in Section~\ref{sec:committee}, we can use any type of diarization system as a committee member.
To further improve the quality of the composite pseudo-label, we also utilize the pseudo-label generated from the clustering-based method.
We demonstrate the effectiveness of this {\it strong pseudo label} by analyzing the results of our successful system submitted to the DIHARD III challenge~\cite{ryant2020third}.
This system was constructed from five subsystems:
two clustering-based systems, two EEND-based systems, and one hybrid system.
For more details about the DIHARD III system description, we refer readers to~\cite{horiguchi2021hitachi}.
In this section, we discuss the results focusing on EEND-EDA~\cite{horiguchi2020end}, which was one of the subsystems.
EEND-EDA was also incorporated into the hybrid system.

As with SC-EEND mentioned above, first, EEND-EDA was trained on a simulated training set, then fine-tuned on a labeled development set (`Supervised EEND-EDA').
Next, the results of all subsystems were combined via DOVER-Lap to obtain the strong pseudo-labels (`First fusion').
Then, we redid the fine-tuning step for EEND-EDA on the concatenation of the development set with the ground truth labels and the evaluation set with the pseudo labels (`Semi-supervised EEND-EDA').
Finally, we updated the results on the EEND-EDA-based subsystem and the hybrid subsystem, and then redid the combination step (`Second fusion').

The DER result is shown in Table~\ref{tab:result_DH}.
The semi-supervised adaptation with the strong pseudo-label for EEND-EDA achieved 6.5\% relative DER reduction compared to supervised adaptation.
Especially, CF errors in DER breakdown were significantly reduced.
However, semi-supervised adaptation did not improve diarization performance over the pseudo-label.
By redoing the combination step including the updated EEND-EDA, we outperformed the first fusion by 0.87\% absolute DER.
This result showed that the strong pseudo-label generated from several types of the diarization system contributes the performance improvement.

\section{Conclusions}
In this paper, we presented a pseudo-label-based training approach to utilize unlabeled data for end-to-end neural diarization.
In our experiments on the CALLHOME dataset, we confirmed that the iterative pseudo-label method gradually improves the DER using only unlabeled data of the target condition.
Moreover, we found that the committee-based training using unlabeled data provides complementary information to labeled data.
Furthermore, by analyzing the results on the DIHARD III challenge, we showed the effectiveness of the pseudo-label generated from the clustering-based and EEND-based systems.
In the future, we will investigate our proposed method for other EEND-based models.

\section{Acknowledgements}
We thank Desh Raj, Zili Huang, Sanjeev Khudanpur, Nelson Yalta, and Yawen Xue for your contribution to the evaluation on the DIHARD III.

\bibliographystyle{IEEEtran}

\bibliography{ref}

\begin{thebibliography}{10}
\providecommand{\url}[1]{#1}
\csname url@samestyle\endcsname
\providecommand{\newblock}{\relax}
\providecommand{\bibinfo}[2]{#2}
\providecommand{\BIBentrySTDinterwordspacing}{\spaceskip=0pt\relax}
\providecommand{\BIBentryALTinterwordstretchfactor}{4}
\providecommand{\BIBentryALTinterwordspacing}{\spaceskip=\fontdimen2\font plus
\BIBentryALTinterwordstretchfactor\fontdimen3\font minus
  \fontdimen4\font\relax}
\providecommand{\BIBforeignlanguage}[2]{{%
\expandafter\ifx\csname l@#1\endcsname\relax
\typeout{** WARNING: IEEEtran.bst: No hyphenation pattern has been}%
\typeout{** loaded for the language `#1'. Using the pattern for}%
\typeout{** the default language instead.}%
\else
\language=\csname l@#1\endcsname
\fi
#2}}
\providecommand{\BIBdecl}{\relax}
\BIBdecl

\bibitem{Tranter2006}
S.~E. Tranter and D.~A. Reynolds, ``An overview of automatic speaker
  diarization systems,'' \emph{IEEE Trans. on ASLP}, vol.~14, no.~5, pp.
  1557--1565, 2006.

\bibitem{Anguera2012}
X.~Anguera, S.~Bozonnet, N.~W.~D. Evans, C.~Fredouille, G.~Friedland, and
  O.~Vinyals, ``Speaker diarization: A review of recent research,'' \emph{IEEE
  Trans. on ASLP}, vol.~20, no.~2, pp. 356--370, 2012.

\bibitem{conf/asru/ZorilaBDH19}
C.~Zorila, C.~Böddeker, R.~Doddipatla, and R.~Haeb-Umbach, ``An investigation
  into the effectiveness of enhancement in {ASR} training and test for
  {CHiME}-5 dinner party transcription,'' in \emph{Proc. ASRU}, 2019, pp.
  47--53.

\bibitem{conf/asru/KandaHFXNW19}
N.~Kanda, S.~Horiguchi, Y.~Fujita, Y.~Xue, K.~Nagamatsu, and S.~Watanabe,
  ``Simultaneous speech recognition and speaker diarization for monaural
  dialogue recordings with target-speaker acoustic models,'' in \emph{Proc.
  ASRU}, 2019, pp. 31--38.

\bibitem{conf/icassp/SnyderGSPK18}
D.~Snyder, D.~Garcia-Romero, G.~Sell, D.~Povey, and S.~Khudanpur, ``X-vectors:
  Robust {DNN} embeddings for speaker recognition,'' in \emph{Proc. ICASSP},
  2018, pp. 5329--5333.

\bibitem{conf/icassp/WanWPL18}
L.~Wan, Q.~Wang, A.~Papir, and I.~Lopez-Moreno, ``Generalized end-to-end loss
  for speaker verification,'' in \emph{Proc. ICASSP}, 2018, pp. 4879--4883.

\bibitem{conf/slt/SellG14}
G.~Sell and D.~Garcia-Romero, ``Speaker diarization with {PLDA} i-vector
  scoring and unsupervised calibration,'' in \emph{Proc. SLT}, 2014, pp.
  413--417.

\bibitem{conf/interspeech/DimitriadisF17}
D.~Dimitriadis and P.~Fousek, ``Developing on-line speaker diarization
  system,'' in \emph{Proc. INTERSPEECH}, 2017, pp. 2739--2743.

\bibitem{conf/icassp/WangDWMM18}
Q.~Wang, C.~Downey, L.~Wan, P.~A. Mansfield, and I.~Lopez-Moreno, ``Speaker
  diarization with {LSTM},'' in \emph{Proc. ICASSP}, 2018, pp. 5239--5243.

\bibitem{conf/interspeech/LinYLBB19}
Q.~Lin, R.~Yin, M.~Li, H.~Bredin, and C.~Barras, ``{LSTM} based similarity
  measurement with spectral clustering for speaker diarization,'' in
  \emph{Proc. INTERSPEECH}, 2019, pp. 366--370.

\bibitem{Frey07AffinityPropagation}
B.~J.~J. Frey and D.~Dueck, ``Clustering by passing messages between data
  points,'' \emph{Science}, vol. 315, no. 5814, pp. 972--976, 2007.

\bibitem{li2019discriminative}
Q.~Li, F.~L. Kreyssig, C.~Zhang, and P.~C. Woodland, ``Discriminative neural
  clustering for speaker diarisation,'' \emph{arXiv preprint arXiv:1910.09703},
  2019.

\bibitem{conf/icassp/ZhangWZP019}
A.~Zhang, Q.~Wang, Z.~Zhu, J.~W. Paisley, and C.~Wang, ``Fully supervised
  speaker diarization,'' in \emph{Proc. ICASSP}, 2019, pp. 6301--6305.

\bibitem{conf/interspeech/FujitaKHNW19}
Y.~Fujita, N.~Kanda, S.~Horiguchi, K.~Nagamatsu, and S.~Watanabe, ``End-to-end
  neural speaker diarization with permutation-free objectives,'' in \emph{Proc.
  INTERSPEECH}, 2019, pp. 4300--4304.

\bibitem{conf/asru/FujitaKHXNW19}
Y.~Fujita, N.~Kanda, S.~Horiguchi, Y.~Xue, K.~Nagamatsu, and S.~Watanabe,
  ``End-to-end neural speaker diarization with self-attention,'' in \emph{Proc.
  ASRU}, 2019, pp. 296--303.

\bibitem{journals/corr/abs-2006-01796}
Y.~Fujita, S.~Watanabe, S.~Horiguchi, Y.~Xue, J.~Shi, and K.~Nagamatsu,
  ``Neural speaker diarization with speaker-wise chain rule,'' \emph{arXiv
  preprint arXiv:2006.01796}, 2020.

\bibitem{horiguchi2020end}
S.~Horiguchi, Y.~Fujita, S.~Watanabe, Y.~Xue, and K.~Nagamatsu, ``End-to-end
  speaker diarization for an unknown number of speakers with encoder-decoder
  based attractors,'' \emph{Proc. INTERSPEECH}, pp. 269--273, 2020.

\bibitem{xue2020online}
Y.~{Xue}, S.~{Horiguchi}, Y.~{Fujita}, S.~{Watanabe}, P.~{Garc\' ia}, and
  K.~{Nagamatsu}, ``Online end-to-end neural diarization with speaker-tracing
  buffer,'' in \emph{Proc. SLT}, 2021, pp. 841--848.

\bibitem{journals/corr/abs-2011-02678}
E.~Han, C.~Lee, and A.~Stolcke, ``{BW-EDA-EEND}: {S}treaming end-to-end neural
  speaker diarization for a variable number of speakers,'' \emph{arXiv preprint
  arXiv:2011.02678}, 2020.

\bibitem{9383555}
Y.~{Takashima}, Y.~{Fujita}, S.~{Watanabe}, S.~{Horiguchi}, P.~{García}, and
  K.~{Nagamatsu}, ``End-to-end speaker diarization conditioned on speech
  activity and overlap detection,'' in \emph{Proc. SLT}, 2021, pp. 849--856.

\bibitem{pseudo2013label}
D.-H. Lee, ``Pseudo-label : {T}he simple and efficient semi-supervised learning
  method for deep neural networks,'' \emph{Proc. ICML 2013 Workshop :
  Challenges in Representation Learning (WREPL)}, 2013.

\bibitem{conf/icassp/Kahn0H20}
J.~Kahn, A.~Lee, and A.~Hannun, ``Self-training for end-to-end speech
  recognition,'' in \emph{Proc. ICASSP}, 2020, pp. 7084--7088.

\bibitem{conf/interspeech/KandaHLK16}
N.~Kanda, S.~Harada, X.~Lu, and H.~Kawai, ``Investigation of semi-supervised
  acoustic model training based on the committee of heterogeneous neural
  networks,'' in \emph{INTERSPEECH}, 2016, pp. 1325--1329.

\bibitem{callhome}
``2000 {NIST} speaker recognition evaluation,''
  https://catalog.ldc.upenn.edu/LDC2001S97.

\bibitem{ryant2020third}
N.~Ryant, P.~Singh, V.~Krishnamohan, R.~Varma, K.~Church, C.~Cieri, J.~Du,
  S.~Ganapathy, and M.~Liberman, ``The third {DIHARD} diarization challenge,''
  \emph{arXiv preprint arXiv:2012.01477}, 2020.

\bibitem{journals/corr/abs-2011-01997}
D.~Raj, L.~P. Garcia-Perera, Z.~Huang, S.~Watanabe, D.~Povey, A.~Stolcke, and
  S.~Khudanpur, ``{DOVER-Lap}: {A} method for combining overlap-aware
  diarization outputs,'' in \emph{Proc. SLT}, 2020, p. 881–888.

\bibitem{horiguchi2021hitachi}
S.~Horiguchi, N.~Yalta, P.~Garcia, Y.~Takashima, Y.~Xue, D.~Raj, Z.~Huang,
  Y.~Fujita, S.~Watanabe, and S.~Khudanpur, ``The {Hitachi-JHU DIHARD III
  System}: {C}ompetitive end-to-end neural diarization and x-vector clustering
  systems combined by {DOVER-Lap},'' \emph{arXiv preprint arXiv:2102.01363},
  2021.

\bibitem{conf/icassp/KinoshitaDAN20}
K.~Kinoshita, M.~Delcroix, S.~Araki, and T.~Nakatani, ``Tackling real noisy
  reverberant meetings with all-neural source separation, counting, and
  diarization system,'' in \emph{Proc. ICASSP}, 2020, pp. 381--385.

\bibitem{medennikov2020target}
I.~Medennikov, M.~Korenevsky, T.~Prisyach, Y.~Khokhlov, M.~Korenevskaya,
  I.~Sorokin, T.~Timofeeva, A.~Mitrofanov, A.~Andrusenko, I.~Podluzhny
  \emph{et~al.}, ``Target-speaker voice activity detection: {A} novel approach
  for multi-speaker diarization in a dinner party scenario,'' in \emph{Proc.
  INTERSPEECH}, 2020.

\bibitem{LAMEL2002115}
L.~Lamel, J.-L. Gauvain, and G.~Adda, ``Lightly supervised and unsupervised
  acoustic model training,'' \emph{Computer Speech \& Language}, vol.~16,
  no.~1, pp. 115--129, 2002.

\bibitem{conf/interspeech/XuSSL14}
H.~Xu, H.~Su, C.~E. Siong, and H.~Li, ``Semi-supervised training for
  bottle-neck feature based {DNN-HMM} hybrid systems,'' in \emph{Proc.
  INTERSPEECH}, 2014, pp. 2078--2082.

\bibitem{conf/icassp/Charlet01}
D.~Charlet, ``Confidence-measure-driven unsupervised incremental adaptation for
  {HMM}-based speech recognition,'' in \emph{Proc. ICASSP}, 2001, pp. 357--360.

\bibitem{conf/asru/VeselyHB13}
K.~Veselý, M.~Hannemann, and L.~Burget, ``Semi-supervised training of deep
  neural networks,'' in \emph{Proc. ASRU}, 2013, pp. 267--272.

\bibitem{conf/interspeech/XuLKHSC20}
Q.~Xu, T.~Likhomanenko, J.~Kahn, A.~Hannun, G.~Synnaeve, and R.~Collobert,
  ``Iterative pseudo-labeling for speech recognition,'' in \emph{Proc.
  INTERSPEECH}, 2020, pp. 1006--1010.

\bibitem{cai2020iterative}
D.~Cai, W.~Wang, and M.~Li, ``An iterative framework for self-supervised deep
  speaker representation learning,'' \emph{arXiv preprint arXiv:2010.1475i},
  2020.

\bibitem{conf/icassp/HamanakaSFEK10}
Y.~Hamanaka, K.~Shinoda, S.~Furui, T.~Emori, and T.~Koshinaka, ``Speech
  modeling based on committee-based active learning,'' in \emph{Proc. ICASSP},
  2010, pp. 4350--4353.

\bibitem{conf/interspeech/HuangYGL13}
Y.~Huang, D.~Yu, Y.~Gong, and C.~Liu, ``Semi-supervised {GMM} and {DNN}
  acoustic model training with multi-system combination and confidence
  re-calibration,'' in \emph{Proc. INTERSPEECH}, 2013, pp. 2360--2364.

\bibitem{conf/icassp/YuKT017}
D.~Yu, M.~Kolbæk, Z.-H. Tan, and J.~Jensen, ``Permutation invariant training
  of deep models for speaker-independent multi-talker speech separation,'' in
  \emph{Proc. ICASSP}, 2017, pp. 241--245.

\bibitem{conf/icassp/HersheyCRW16}
J.~R. Hershey, Z.~Chen, J.~L. Roux, and S.~Watanabe, ``Deep clustering:
  {D}iscriminative embeddings for segmentation and separation,'' in \emph{Proc.
  ICASSP}, 2016, pp. 31--35.

\bibitem{conf/interspeech/McCreeSG19}
A.~McCree, G.~Sell, and D.~Garcia-Romero, ``Speaker diarization using
  leave-one-out gaussian {PLDA} clustering of {DNN} embeddings,'' in
  \emph{Proc. INTERSPEECH}, 2019, pp. 381--385.

\end{thebibliography}

\end{document}